%% file: main.tex
\documentclass[10pt,a4paper]{article}
\usepackage[utf8]{inputenc}
\usepackage[T1]{fontenc}
\usepackage[a4paper,margin=0.75in]{geometry}
\usepackage{authblk}
\usepackage{amsmath,amsfonts, amsthm}
\usepackage{amssymb}
\usepackage{algorithmic}
\usepackage{algorithm}
\usepackage{array}
\usepackage[caption=false,font=normalsize,labelfont=sf,textfont=sf]{subfig}
\usepackage{textcomp}
\usepackage{verbatim}
\usepackage{graphicx}
\usepackage{cite}
\usepackage{hyperref}
\usepackage{tikz-cd}
\usetikzlibrary{arrows.meta,positioning,calc,backgrounds,decorations.markings}
\usepackage{pgfplots}
\pgfplotsset{compat=1.18}

\usepackage{siunitx}
\definecolor{mypink}{HTML}{EF7B6B}
\definecolor{myblue}{HTML}{35A2D4}
\usepackage{mdframed}
\usepackage{tabularx}
\usepackage{booktabs}
\usepackage{printlen}
\usepackage{placeins}
\uselengthunit{in} % or mm, cm, pt
\providecommand{\makecell}[2][]{\shortstack{#2}}
\newcommand{\keywords}[1]{\par\noindent\textbf{Keywords:} #1}
\newcommand{\vect}[1]{\boldsymbol{#1}}

\newcommand{\ptarg}{P_{\text{target}}}

\newcommand{\R}{\mathbb{R}}
\newcommand{\1}{\mathbf 1}

\usepackage{cleveref}
\theoremstyle{definition}
\newtheorem{proposition}{Proposition}
\crefname{proposition}{Proposition}{Propositions}
\newtheorem{definition}{Definition}
\newtheorem{remark}{Remark}

\newcounter{appendix}
\renewcommand{\theappendix}{\Alph{appendix}}

\newenvironment{myproof}[1][Proof]{%
  \par\noindent\textbf{#1.}\space%
}{%
  \hfill$\blacksquare$\par%
}

\newcommand\blfootnote[1]{%
  \begingroup
  \renewcommand\thefootnote{}\footnotetext{#1}%
  \endgroup
}

\begin{document}

\title{Proportional dispatch and fairness in wind farm power tracking}

\author[1,2]{Baptiste Corban}
\author[1]{Ana Bušić}
\author[2]{Donatien Dubuc}
\author[2]{Jiamin Zhu}
\affil[1]{Inria and DI ENS, Ecole Normale Supérieure, PSL Research University, Paris, France}
\affil[2]{IFP Energies nouvelles, Rueil-Malmaison, France}
\affil[ ]{\texttt{baptiste.corban@ifpen.fr}}
\date{}

\maketitle

\blfootnote{This work has been submitted to the IEEE for possible publication. Copyright may be transferred without notice, after which this version may no longer be accessible.}

\begin{abstract}
Controlling the power output of a wind farm to track a target signal enables contributing to the power grid stability. 
This can be achieved by dividing the target signal into individual turbine power setpoints, which are then tracked by the corresponding turbine controllers. 
In this work, we address the problem of finding power allocations that fairly spread the power reserves (i.e. the unused fraction of available powers) among turbines, thereby improving robustness to uncertainties and varying wind conditions.
Specifically, we investigate the fairness properties of proportional dispatch, the most widely used power allocation strategy.
We show that, because of wake interactions within the wind farm, proportional dispatch must be applied iteratively to achieve fair distribution of power reserves. 
We study the convergence of the iterative proportional dispatch (IPD) process to equalized reserves, and then illustrate it using both steady-state and dynamic wind farm simulators.
The numerical results show that IPD closely approximates max-min fairness, a related fairness criterion, while requiring significantly less computational effort than black-box optimization.
Finally, we show that IPD also reduces the complexity of the problem of fair power dispatch combined with yaw wake steering optimization.

\end{abstract} 


\section{Introduction}
\input{introduction.tex}

\section{Wind farm power dispatch problem}
\label{sec:problem}
\input{problem.tex}
\section{Max-min fairness in the independent case}
\label{sec:independent}
\input{independent.tex}
\section{Iterative proportional dispatch for the wake-coupled case}
\label{sec:coupled}
\input{coupled.tex}
\section{Simulated examples of iterative proportional dispatching}
\label{sec:numerical}
\input{numerical.tex}

\section{Yaw-augmented power dispatching}
\label{sec:yaw-augmented}
\input{yaw_augmented.tex}
% \section{Relaxed iteration}
\section{Conclusion}
\input{conclusion.tex}

\section*{Acknowledgment}
This work is partially carried out in the framework of AI-NRGY (Grant ref.: ANR-22-PETA-0004).

\phantomsection\refstepcounter{appendix}\label{appendix:proof_independent}
\section*{Appendix \theappendix: Proof of Proposition~\ref{prop:independent}}
\input{appendixA.tex}
\phantomsection\refstepcounter{appendix}\label{appendix:proof_coupled}
\section*{Appendix \theappendix: Proof of Proposition~\ref{prop:coupled}}
\input{appendixB.tex}

\bibliographystyle{unsrt}

\bibliography{biblio}

\end{document}

%% file: introduction.tex
% !TEX root = main.tex
The frequency stability of the power grid is crucial to its operation, and results from a balance between generation and consumption of electricity.
Whenever there is a mismatch between supply and demand of electricity, the frequency of the grid deviates from its nominal value (\SI{50}{\hertz} in Europe). If the frequency deviation is not contained, it can lead to blackouts and damages.
To prevent this, the grid relies on frequency regulation services, provided by generators which can adjust their power output in response to the grid frequency deviations.
As the share of renewable energy sources such as wind and solar grows, they are increasingly expected to contribute to grid stability.
Thus, wind farms are required to be able to track a power reference signal provided by the grid operator.

A target power output for a wind farm can be divided into individual power setpoints for each turbine in the farm.
Because the farm power target is generally lower than the maximum power the wind farm can produce, there might be multiple ways to distribute it into individual turbine setpoints.
However, different wind conditions throughout the wind farm (which can span several kilometers), and wakes generated behind the turbines, can lead to differences in available wind power and fatigue states between the turbines. 
Therefore, different power allocations may lead to different performance in terms of tracking the power reference \cite{diaz-sanahujaEnhancingOffshoreWind2026,linHierarchicalClusteringbasedOptimization2020,liWindFarmActive2019} or structural loads on the turbines for example \cite{flemingComputationalFluidDynamics2016,biegelDistributedLowcomplexityController2013,kongDistributedOptimalControl2025,yaoOptimizedActivePower2023,zhaoFatigueLoadSensitivityBased2017,zhaoActivePowerControl2021}.
The power dispatch problem consists of finding the best way to distribute the farm power target into individual turbine setpoints, in order to achieve a certain objective.

\begin{figure}[htbp]
    \centering
    \resizebox{\linewidth}{!}{\input{dispatch.tex}}
    \caption{General framework of power dispatch for wind farm frequency regulation}
    \label{fig:dispatch}
\end{figure}
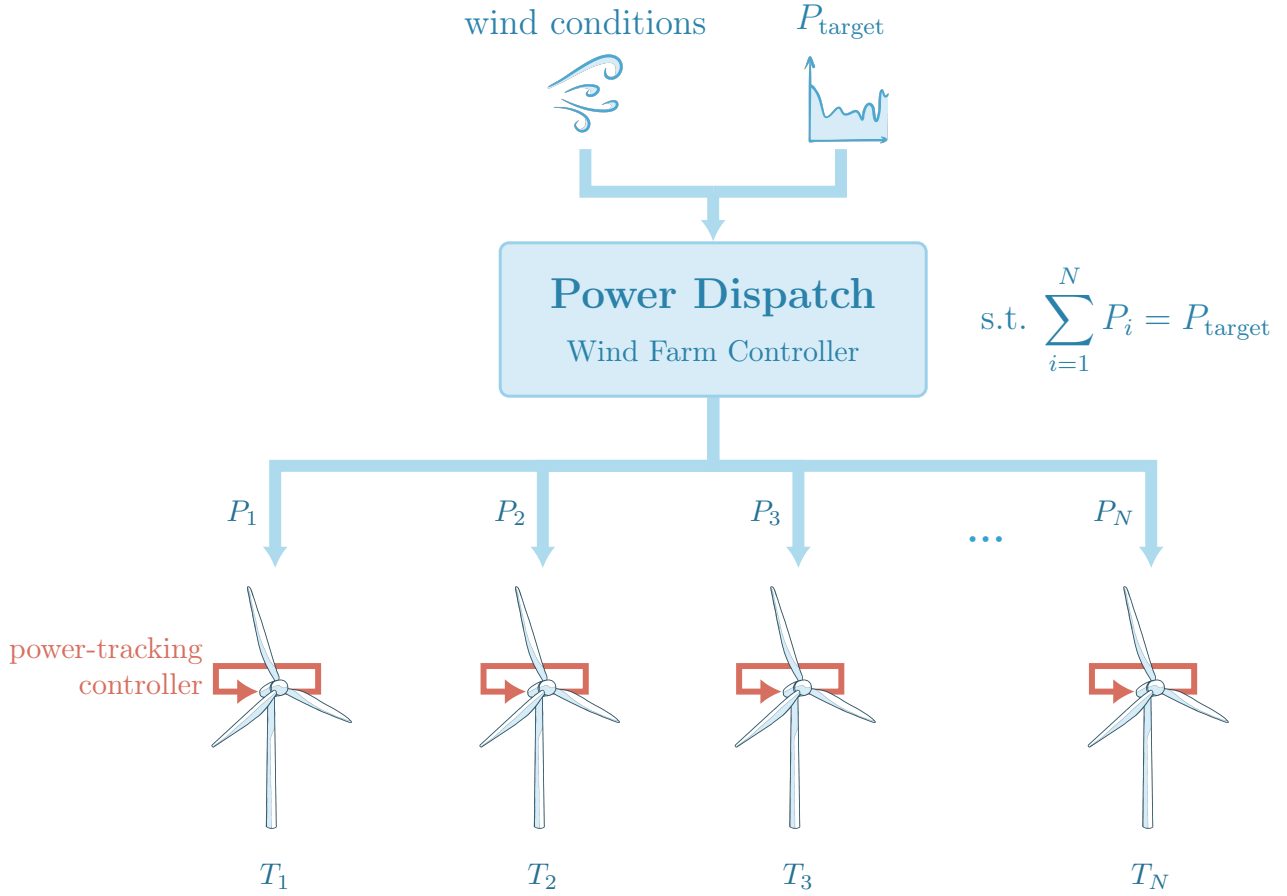

\autoref{fig:dispatch} illustrates the centralized power dispatch framework. A wind farm centralized controller receives a power target $P_{\text{target}}$ from the grid operator and the current wind conditions. It then computes power setpoints $P_{i}$ for each turbine $T_i$ in the farm, such that the total power produced by the farm meets the target, i.e. $\sum_{i=1}^N P_i = \ptarg$. 
The turbines then track these setpoints using their local blade pitch and generator torque controllers.

A desirable property for a power dispatch method is to be robust to errors in the estimation of a turbine's available power, which can arise from wind fluctuations or model uncertainties.
To achieve this, it is beneficial to have the turbines not operate at their estimated maximum available power~\cite{tamaroRobustActivePower2025a,siniscalchi-minnaWindFarmControl2019,tianActivePowerDispatch2014}. This ensures that even if the available power is over- or underestimated, the turbines remain capable of reaching their respective power setpoints.
This can be quantified by the power reserve, which represents the percentage of available power that is not used by a turbine.
The power dispatch problem can then be seen as a resource allocation problem: the resource to allocate is the power production, and the utility of a turbine is the power reserve, which we want to maximize.

Resource allocation problems have been studied in various fields such as economics~\cite{youngEquityTheoryPractice1995}, social sciences, communication networks\cite{bertsekasDataNetworks1992,kellyRateControlCommunication}, finance, etc.
When a centralized decision maker is responsible for allocating a resource among multiple agents, the utilitarian approach would consist in maximizing the sum of the agents' utilities~\cite{bertsimasPriceFairness2011,kellyRateControlCommunication}. 
However, this approach can lead to very unfair allocations, where some agents have a very low utility or allocated resource. Therefore, the notion of fairness in allocation naturally arises, with different possible definitions~\cite{bertsekasDataNetworks1992,lussEquitableResourceAllocation2012}.

The max-min fairness consists in maximizing the utility of the worst-off agent~\cite{ogryczakFairOptimizationNetworks2014}, and is commonly used in communication networks.
For wind farm power dispatch, max-min fairness of power reserves has been chosen as objective in  \cite{tamaroRobustActivePower2025a}. The authors solved a constrained optimization problem offline for a variety of wind conditions using a gradient-based sequential quadratic programming (SQP) method. This generates a power-setpoint look up table, which is then interpolated at runtime. This problem can be costly to solve for large wind farms: it is non-convex due to the complex wake interactions, and high-dimensional due to the number of turbines.
Alternatively, in~\cite{barosDistributedTorqueControl2017, fanOptimizedDecentralizedPower2021, kongDistributedOptimalControl2025, zhangFullyDistributedCoordination2013, dongFullyDistributedDeloadingOperation2021}, another fairness definition used as dispatch objective is \textit{fair load sharing}, which aims to equalize the power reserves of all turbines.

For power dispatch, a common baseline method is to allocate power setpoints in proportion to the available power of each turbine as in \cite{hansenCentralisedPowerControl2006,kazdaModelOptimizedDispatchClosedLoop2020,yaoOptimizedActivePower2023,sorensenAerodynamicAnalysisWind2012,merahiNewManagementStructure2014,fernandezComparativeStudyPerformance2008}. Furthermore, this strategy, referred to as \textit{proportional dispatch} (PD) in this paper, has been said to achieve fair load sharing in \cite{barosDistributedTorqueControl2017,fanOptimizedDecentralizedPower2021}.
In this article, we aim at providing a theoretical study of the fairness properties of proportional dispatch, to justify its use as the baseline method for power dispatch in wind farms.
We highlight that the wake interactions between turbines can break the fair load sharing property of PD. 
If those interactions are ignored, we formally demonstrate in this paper that the PD strategy not only achieves fair load sharing but also max-min fairness. This constitutes the first contribution of this paper, which is detailed in \autoref{sec:independent}.

Nevertheless, in reality this assumption is not satisfied: the available power of a turbine depends on its local wind conditions, which are influenced by the operating points of upstream turbines. Therefore, due to wake effects, changing the power setpoint of a turbine can change the available power of other turbines in the farm. Consequently, any power allocation method relying on local wind measurements, such as PD, must be updated iteratively, even under steady free-wind conditions.
As a second contribution, we show in \autoref{sec:coupled} that an iterative proportional dispatch (IPD) converges and achieves fair load sharing, under sufficient conditions.
Though in this case we do not have proof that fair load sharing resulting from IPD corresponds to an optimal solution of the max-min fairness problem, we illustrate on several examples in~\autoref{sec:numerical} that it is a very close approximation, at a considerably lower computational cost than traditional optimization methods.
Finally, we show numerically in~\autoref{sec:yaw-augmented} that IPD also substantially  simplifies the joint yaw-power dispatch problem considered in~\cite{tamaroRobustActivePower2025a}.

To the best of our knowledge, this work is the first to provide a study of the fairness properties of proportional dispatch with the influence of wake effects. 
We also provide IPD as a practical, low-cost alternative to solving the max-min fairness dispatch problem in wind farms, well suited to real-time applications.

\autoref{tab:dispatch_symbols} summarizes the main symbols used throughout this article.

\begin{table}[htbp]
\centering
\vspace{2pt}
\renewcommand{\arraystretch}{1.2}
\begin{tabular}{@{}lp{0.65\linewidth}@{}}
\toprule
\textbf{Symbol} & \textbf{Description} \\
\midrule
PD & Proportional Dispatch \\
IPD & Iterative Proportional Dispatch \\
$\ptarg$ & Farm-level power target from the grid operator \\
$P_i$ & Power setpoint allocated to turbine $i$ \\
$x_i$ & Normalized dispatch coefficient for turbine $i$: $P_i = x_i \cdot \ptarg$ \\
$\vect{x} = (x_i)_{i=1}^N$ & Dispatch vector \\
$\Delta$ & $N$-simplex $\{x \in \R^N : x_i \geq 0,\ \textstyle\sum_{i=1}^N x_i = 1\}$ \\
$P_{a,i}$ & Available power of turbine $i$ \\
$\vect{P_a}$ & Vector of available powers $(P_{a,i})_{i=1}^N$ \\
$r_i$ & Power reserve of turbine $i$: $r_i = 1 - P_i / P_{a,i}$ \\
$g$ & Proportional dispatch operator, $g_i(x) = \frac{P_{a,i}(x)}{\sum_j P_{a,j}(x)}$ \\
$x^k$ & Dispatch vector at IPD iteration $k$ \\
$x^\star$ & Fixed point of IPD (converged dispatch) \\
$D_{\mathrm{KL}}(x,\, y)$ & Kullback-Leibler divergence\\
QBFNE & Quasi-Bregman firmly nonexpansive operator \\
\bottomrule
\end{tabular}
\vspace{0.2cm}
\caption{Main symbols used in this article.}
\label{tab:dispatch_symbols}
\end{table}

%% file: dispatch.tex
\begin{tikzpicture}[
    box/.style={rectangle, minimum width=3.5cm, minimum height=1.2cm, align=center, fill=#1, rounded corners=2pt},
    controller/.style={rectangle, minimum width=5cm, minimum height=1.8cm, align=center, fill=myblue!20, rounded corners=3pt, draw = myblue!50, line width=1.pt},
    arrow/.style={-{Latex[length=2.5mm, width=3mm]}, line width=4pt, color=myblue!40},
]

\node[controller] (controller) at (0, 0) {\textcolor{myblue!80!black}{\Large\textbf{Power Dispatch}}\\[0.2cm]\textcolor{myblue!80!black}{\normalsize Wind Farm Controller}};

\node (wind_text) at (-1.5, 3.5) {\textcolor{myblue!80!black}{\large wind conditions}};
\node[below=-0.3cm of wind_text] (wind_img) {\includegraphics[width=1.5cm]{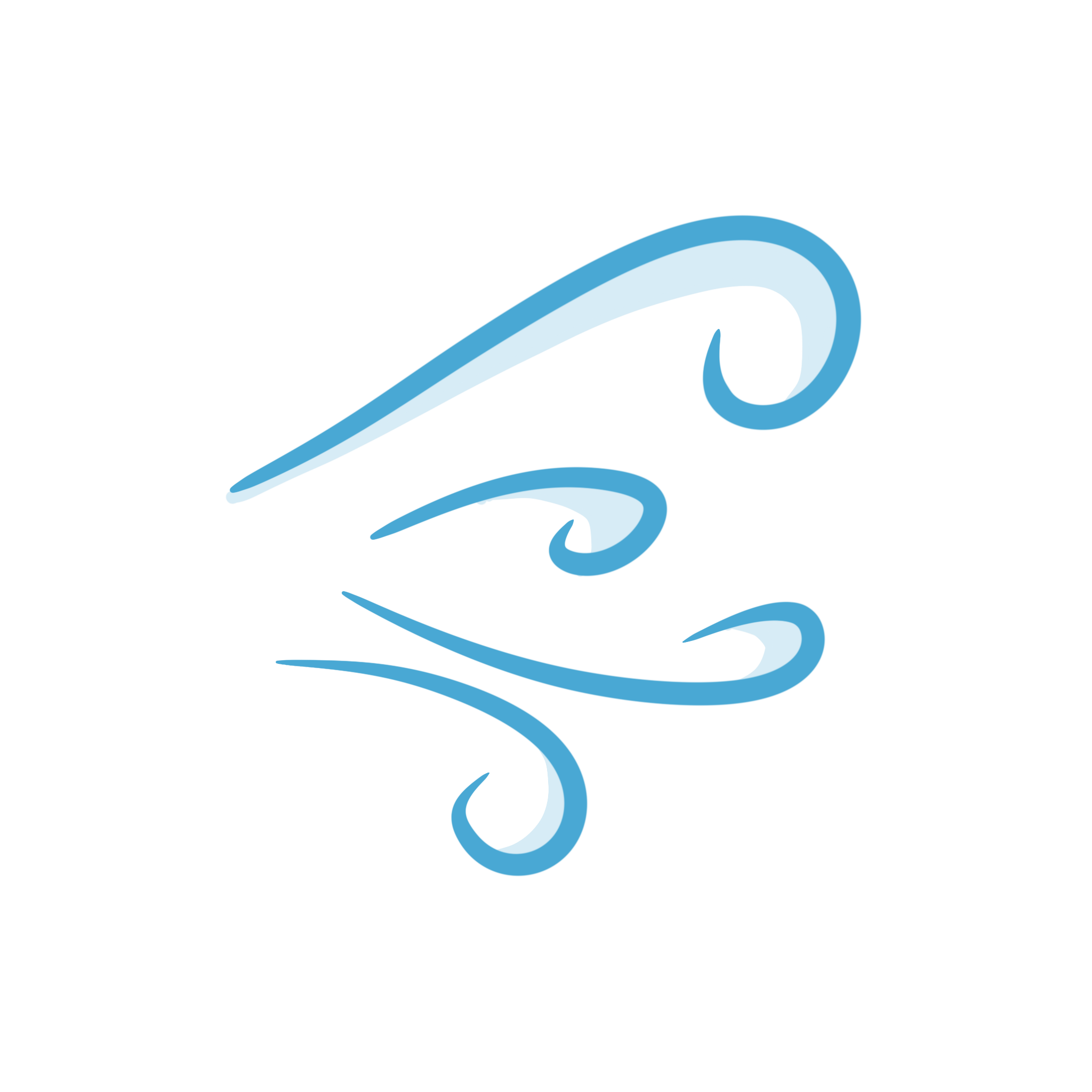}};
\node (ptarg_text) at (1.5, 3.5) {\textcolor{myblue!80!black}{ \textbf{\large $P_{\mathrm{target}}$}}};
\node[below=-0.3cm of ptarg_text] (ptarg_img) {\includegraphics[width=1.5cm]{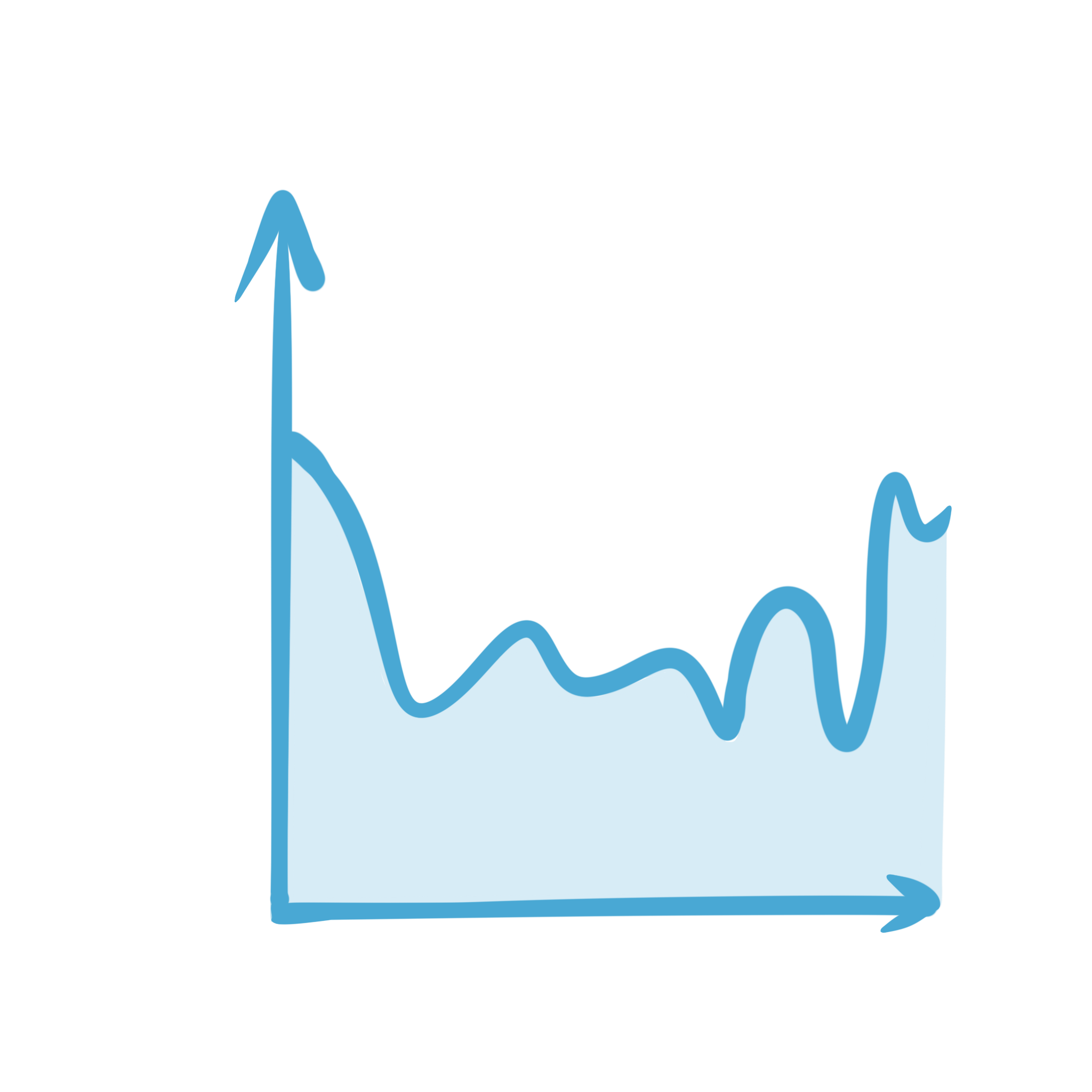}};
\draw[-, line width=4pt, color=myblue!40] (-1.5,2)-- (-1.5,1.5) -| (0,1.5);
\draw[-, line width=4pt, color=myblue!40] (1.5,2)-- (1.5,1.5) -| (0,1.5);
\draw[arrow] (0,1.5)-- (controller.north);

\node[below left=2cm and 1cm of controller] (wt1) {\includegraphics[width=3cm]{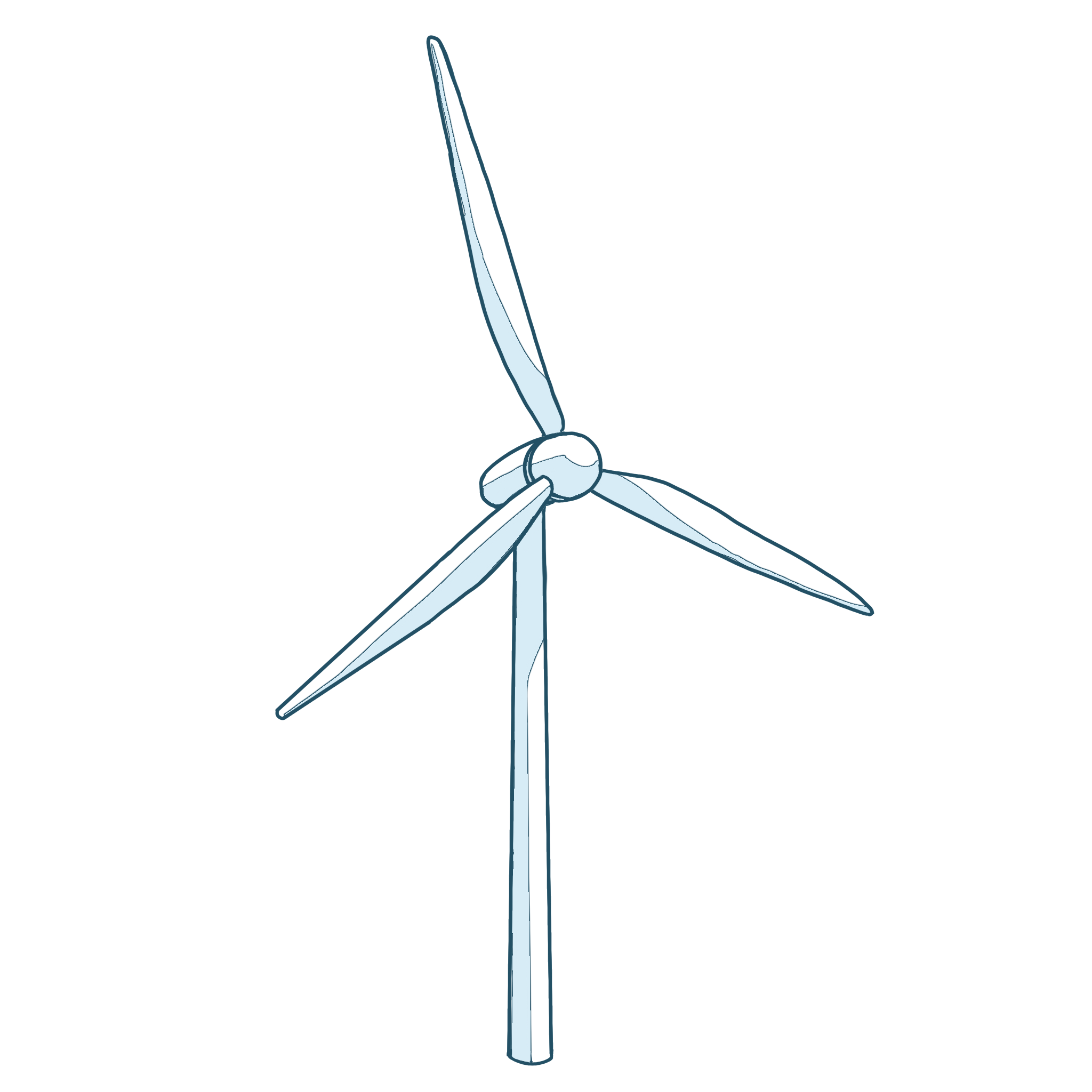}};
\node[below=2cm of controller, xshift=-2cm] (wt2) {\includegraphics[width=3cm]{turbine.png}};
\node[below=2cm of controller, xshift=1cm] (wt3) {\includegraphics[width=3cm]{turbine.png}};
\node[below right=2cm and 1cm of controller] (wtn) {\includegraphics[width=3cm]{turbine.png}};

\node[below=0.1cm of wt1,] {\textcolor{myblue!70!black}{\textbf{$T_1$}}};
\node[below=0.1cm of wt2,] {\textcolor{myblue!70!black}{\textbf{$T_2$}}};
\node[below=0.1cm of wt3,] {\textcolor{myblue!70!black}{\textbf{$T_3$}}};
\node[below=0.1cm of wtn,] {\textcolor{myblue!70!black}{\textbf{$T_N$}}};

\node[below=1.5cm of controller, xshift= 3.2cm, font=\Large] (dots) {\textcolor{myblue}{\textbf{...}}};

\begin{scope}[on background layer]
\draw[arrow=myblue!30] (controller.south) -- ++(0,-0.8cm) -| (wt1.north) node[midway, left, yshift=-5.5mm, text=myblue!70!black] {$P_1$};
\draw[arrow=myblue!30] (controller.south) -- ++(0,-0.8cm) -| (wt2.north) node[midway, left, yshift=-5.5mm, text=myblue!70!black] {$P_2$};
\draw[arrow=myblue!30] (controller.south) -- ++(0,-0.8cm) -| (wt3.north) node[midway, left, yshift=-5.5mm, text=myblue!70!black] {$P_3$};
\draw[arrow=myblue!30] (controller.south) -- ++(0,-0.8cm) -| (wtn.north) node[midway, left, yshift=-5.5mm, text=myblue!70!black] {$P_N$};
\end{scope}

\node[right=0.5cm of controller, font=\large] {\textcolor{myblue!80!black}{s.t. $\displaystyle\sum_{i=1}^{N} P_i = P_{\mathrm{target}}$}};

  \begin{scope}[on background layer]
\draw[arrow, line width=2pt, color=mypink!90!black] ($(wt1.north east)+(-1.62,-1.45)$) -- ++(0.5,0.) -- ++(0.,0.3) -- ++(-1.2,0) node[left, text=mypink!90!black, font=\normalsize, align=right] {power-tracking\\controller} -- ++(0,-0.3) -- ++(0.5,0);
\foreach \wt in {wt2, wt3, wtn} {
    \draw[arrow, line width=2pt, color=mypink!90!black] ($(\wt.north east)+(-1.62,-1.45)$) -- ++(0.5,-0.) -- ++(0.,0.3) -- ++(-1.2,0) -- ++(0,-0.3) -- ++(0.5,0);
}
\end{scope}

\end{tikzpicture}

%% file: problem.tex
% !TEX root = main.tex
\subsection{Power dispatch in a wind farm}
A wind farm is a set of $N$ turbines $(T_i)_{1\leq i \leq N}$. 
When each turbine is operated to maximize its individual power~\cite{bossanyiDesignClosedLoop2000}, the farm is said to be in \textit{greedy control}.
The power produced by the farm is then denoted as the greedy power $P_{\text{greedy}}$.
We consider the case where the wind farm is required to produce a target power $P_{\text{target}}$ lower than the greedy power, i.e. $P_{\text{target}} \leq P_{\text{greedy}}$.

Let us define the dispatch vector $\vect{x}=(x_i)_{i=1}^N \in \Delta$, where $\Delta = {x\in\R^N : x_i\ge0,\ \sum x_i=1}$ is the $N$-simplex representing all possible dispatch vectors.
Then, the power allocated to a turbine $i$ is: 
\begin{equation}
P_i = x_i \cdot P_{\text{target}} \quad s.t. \quad 0 \leq P_i \leq P_{a,i}
\end{equation}
with $P_{a,i}$ the maximum available power for turbine $i$ estimated (for under rated wind speeds \cite{boersmaTutorialControlorientedModeling2017}) through
\begin{equation}\label{eq:Pa_def}
P_{a,i}(x)= \frac{1}{2} \rho A C_{p,\max} u_i(x)^3
\end{equation}
where $\rho$ is the air density, $A$ the rotor swept area, $u_i$ the rotor-averaged wind speed, and $C_{p,\max}$ is the maximum power coefficient for the turbine. 

\begin{remark}
The available power \eqref{eq:Pa_def} depends on the local wind conditions, which can differ across the farm due to wake effects. Indeed, changing the power setpoint of an upstream turbine will affect the local wind-speed of a downstream one. Thus, we represent the available powers and local wind speeds as functions of the current power dispatch $x$.
\end{remark}

For a turbine $i$, we define the power reserve as
\begin{equation} \label{eq:reserve}
r_i(x) = 1 - \frac{x_i \cdot P_{\text{target}}}{P_{a,i}(x)}
\end{equation}
It represents the percentage of unused available power by the turbine. 
It is desirable to operate turbines by maintaining a power reserve, as it improves the robustness of the power tracking~\cite{tamaroRobustActivePower2025a}. Indeed, in case of an imprecise prediction of the available power, due to wind fluctuations or measurement uncertainties, having turbines operate with some reserve improves the ability to meet the individual power setpoint.

\subsection{Proportional dispatch: a common baseline}

A simple and common power dispatch method is proportional dispatch (PD)~\cite{kazdaModelOptimizedDispatchClosedLoop2020,yaoOptimizedActivePower2023,sorensenAerodynamicAnalysisWind2012,merahiNewManagementStructure2014,fernandezComparativeStudyPerformance2008}.
It allocates power setpoints to each turbine based on their available power~\cite{hansenCentralisedPowerControl2006}. In the following, PD is represented by the function $g$ defined as:
\begin{equation} \label{eq:PD_def}
    g(x) = \left(g_i(x)\right)_{1 \leq i \leq N} \quad \text{with} \quad g_i(x) =  \frac{P_{a,i}(x)}{\sum_{j=1}^{N} P_{a,j}(x)}
\end{equation}
As the available powers depend on the current power dispatch, so does the proportional dispatch.

\subsection{Fairness definitions for power dispatch}

The power dispatch objective in this study is to fairly distribute power reserves between turbines, as it helps with robustness of the power tracking.
We are interested in two fairness definitions existing in the literature: max-min fairness and fair load sharing. 

\begin{definition}[Max-min fairness]
\label{def:maxmin}
Max-min fairness aims to maximize the minimum power reserve among all turbines, while ensuring that the total power produced meets the target. It can be formulated as the following constrained optimization problem~\cite{tamaroRobustActivePower2025a}:
\begin{equation}
\begin{aligned}
&\max_{(x_i)_{i=1}^N} \min_{i} r_i(x) \\
&\text{subject to} \quad x \in \Delta, \quad  x_i \leq \frac{P_{a,i}(x)}{P_{\text{target}}} \quad \forall i \\
\end{aligned}
\label{eq:critere_optim}
\end{equation}
\end{definition}

\begin{definition}[Fair load sharing]
\label{def:fairload}
Fair load sharing aims to distribute the power demand such that all turbines operate with identical power reserves. Formally, it seeks a dispatch vector $(x_i)_{i=1}^N$ satisfying
\begin{equation}
\begin{aligned}
&r_i(x) = r_j(x), \quad \forall i,j,\\
&\text{and} \quad x \in \Delta, \quad  x_i \leq \frac{P_{a,i}(x)}{P_{\text{target}}} \quad \forall i \\
\end{aligned}
\end{equation}
\end{definition}

This fairness objective is used for example in~\cite{barosDistributedTorqueControl2017, fanOptimizedDecentralizedPower2021, kongDistributedOptimalControl2025, zhangFullyDistributedCoordination2013, dongFullyDistributedDeloadingOperation2021}.

%% file: independent.tex
% !TEX root = main.tex
In this section, we assume that the available powers of the turbines are independent from the power allocation (wake effects between turbines are absent or considered constant). This means that the dependence of $P_{a,i}$ in $x$ is omitted and \eqref{eq:PD_def} can be applied as :
\begin{equation}  \label{eq:independantPD}
    x_i = g_{i} =  \frac{P_{a,i}}{\sum_{j=1}^{N} P_{a,j}} 
\end{equation}

In this simplified setting, we can easily prove that \eqref{eq:independantPD} is a fair load sharing solution~\cite{barosDistributedTorqueControl2017,fanOptimizedDecentralizedPower2021}, and that it is the optimal solution of the max-min fairness problem (\cref{def:maxmin}).

\begin{proposition}
\label{prop:independent}
If the available power distribution in the wind farm is independent from the power allocation, then, the max-min fairness allocation problem~\eqref{eq:critere_optim} is solved by applying the proportional dispatch \eqref{eq:independantPD}, which equalizes all the reserves \eqref{eq:reserve} of the wind farm.
\end{proposition}

\begin{myproof}
    The full proof is given in \autoref{appendix:proof_independent}.
    It relies on the Hölder inequality, from which we derive a lower bound $r^\star$ on the maximum reserve that can be achieved, and show that this bound is reached by proportional dispatch:
    \begin{equation*}
\begin{aligned}
    r_i  &= 1 - \frac{P_i}{P_{a,i}}  = 1 - \frac{P_{a,i}}{\sum_{j=1}^N P_{a,j}} \cdot \frac{P_{\text{target}}}{P_{a,i}}\\
     &= 1 - \frac{P_{\text{target}}}{\sum_{j=1}^N P_{a,j}}  = r^\star
    \end{aligned}
\end{equation*}
\end{myproof}

In this independent setting, PD achieves both max-min fairness and fair load sharing.

%% file: coupled.tex
% !TEX root = main.tex
Recall that, due to wake effects, the available power of each turbine depends on the current power dispatch in the farm, and will change 
with the power allocation. 
The dispatch problem thus becomes coupled and more complex. 
A natural approach to address this issue is to iteratively apply proportional dispatch, updating the available powers at each iteration, until convergence. We denote this approach as iterative proportional dispatch (IPD).

More precisely, given an initial dispatch $x^0 \in \Delta$, the iterative proportional dispatch (IPD) sequence is defined for all $k \geq 0$ as:
\begin{equation}
x^{k+1} = g(x^k)
\end{equation}
where $g=(g_1,\cdots,g_N)$ with $g_i$ defined by~\eqref{eq:PD_def}.

The power reserve of turbine $i$ at iteration $k\geq1$ is then given as: 
\begin{equation}
r_i^{k} = 1-\frac{P_{a,i}(x^{k-1})}{P_{a,i}(x^{k})} \cdot \frac{\ptarg}{\sum_j P_{a,j}(x^{k-1})}
\end{equation}
It can be seen that the wake updates make the ratio of consecutive available powers different than 1, thus making the reserves unequal across turbines.

In the following, we provide a sufficient condition on the iterative operator $g$ under which the IPD iterates converges to a fixed point of $g$. 
Since dispatch vectors $x$ belong to the simplex $\Delta$ similar to probability distributions, we study the IPD iterates in the Kullback–Leibler (KL) framework.

\begin{definition}[Kullback-Leibler divergence] 
\label{def:KL}
The Kullback-Leibler (KL) divergence~\cite{coverElementsInformationTheory2005} is defined for two vectors $x,y \in \Delta$ as \[ D_{\text{KL}}(x,y) = \sum_{i=1}^N x_i \log \frac{x_i}{y_i}\] 
\end{definition} 

Our convergence proof is based on the quasi-Bregman firmly nonexpansive property of the dispatch operator $g$, which is a standard assumption for establishing the convergence of iterative algorithms to fixed points~\cite{naraghiradBregmanWeakRelatively2013, tomizawaStrongConvergenceTheorem2014}.

\begin{definition}[Quasi-Bregman firmly nonexpansive (QBFNE) operator] 
A mapping $g: \mathbb{R}^N \to \mathbb{R}^N$ is quasi-Bregman firmly nonexpansive (QBFNE) with respect to the KL divergence $D_{\text{KL}}$ if for all $x \in \mathbb{R}^N$, and for all $z \in \text{Fix}(g) = \{ z \in \mathbb{R}^N : g(z) = z \}$, the set of fixed points of $g$,
we have: \[ D_{\text{KL}}(z, g(x)) + D_{\text{KL}}(g(x), x) \leq D_{\text{KL}}(z, x) \]
Or equivalently, 
\[ \langle \nabla \psi(x) - \nabla \psi(g(x)), g(x) - z \rangle \geq 0 \] 
where $\psi$ is the negative entropy function.
\end{definition}

This allows us to formulate the following sufficient condition for the convergence of IPD.

\begin{proposition}[A sufficient condition for convergence of IPD]
\label{prop:coupled}
If the proportional dispatch function $g$ 
is such that for all $x \in \Delta$ and for all fixed points $x^\star$ of $g$:
\begin{equation}
\label{eq:QBFNE_conditions}
\sum_{i=1}^N (x^\star_i - g_i(x)) \log \frac{x_i}{g_i(x)} \leq 0
\end{equation}
Then, $g$ is quasi-Bregman firmly nonexpansive (QBFNE) with respect to the Kullback-Leibler (KL) divergence, and \textbf{IPD converges to a fixed point $x^\star$ of $g$}, i.e. $x^\star = g(x^\star)$.
\end{proposition}

\begin{myproof}
The proof relies on the properties of the KL and Bregman divergence framework, which are detailed in \autoref{appendix:proof_coupled}.
We define the KL divergence between the IPD iterates and a fixed point $x^\star$ of $g$ as
$V^k=D_{\text{KL}}(x^*,x^k)$.
We prove that if \eqref{eq:QBFNE_conditions} is satisfied, then the sequence $\{V^k \}$ is strictly decreasing, and the properties of the KL divergence lead to the convergence of the sequence $\{x^k \}$ to a fixed point of $g$. 
\end{myproof}
\bigskip 

\begin{remark}
\label{remark:2}
Note that \eqref{eq:QBFNE_conditions} is a sufficient condition corresponding to standard assumption of QBFNE in the literature~\cite{naraghiradBregmanWeakRelatively2013, tomizawaStrongConvergenceTheorem2014}. If it is not satisfied, the iterates IPD can still converge if the sequence $\{ V^k \}$ is strictly decreasing, i.e. there exists a $K>0$ such that
\begin{equation} \label{eq:condition_sufficient_necessary}
    V^{k+1} < V^k, \quad \forall k \geq K.
\end{equation}
\end{remark}

Condition~\eqref{eq:QBFNE_conditions} can easily be checked a posteriori for the iterates of IPD.
The iterative dispatch operator $g$ is a function of the available powers. Therefore, this condition inherently concerns the available power function, i.e. this is a condition on the wake interactions.
 
\bigskip 

Next, we show that IPD converges to a fair load sharing dispatch.
  
\begin{proposition}[Fair load sharing via IPD]
  Given a fixed point $x^*$ of the iterative operator $g$, let $P_{a,i}^* = P_{a,i}(x^*)$ be the corresponding available power of turbine $i$. 
Then for all $i \in \{ 1,\dots,N \}$, the power reserve $r_i$ of turbine $i$ is equal to
  
  \begin{equation}
  r^\star= 1 - \frac{ P_{\text{target}}}{\sum_{j=1}^N P_{a,j}^\star}
  \end{equation} 
\end{proposition}

\begin{myproof}
As $P_a$ is a continuous function of $x$, $P_a(x^*)= P_a(g(x^*))$.

Thus, we have

\[r_i  = 1 - \frac{x_i^\star \cdot \ptarg}{P_{a,i}(x^\star)}  = 1 - \frac{\ptarg}{\sum_{j=1}^N P_{a,j}(x^\star)}  = r^\star\]
\end{myproof}

\bigskip 

In this coupled setting, we have shown that IPD can converge to a dispatch that equalizes the power reserves of all turbines, thus achieving fair load sharing.
However, we can no longer theoretically establish equivalence between max-min fairness and fair load sharing. More precisely, the fixed point of $g$ is not necessarily an optimal solution of the max-min problem \eqref{eq:critere_optim} without additional assumptions.

%% file: numerical.tex
% !TEX root = main.tex
In this section, we present numerical experiments demonstrating IPD's convergence and fair load sharing across a variety of layouts and wind conditions.

\subsection{Wind farm models and turbine controllers}

Two simulators are used in this section to simulate the wind farm: FLORIS~\cite{gebraadWindPlantPower2016} and FAST.Farm~\cite{jonkmanDevelopmentFASTFarmNew2017}.

FLORIS is a steady-state simulator, which resolves the steady state of the wind farm for given power setpoints. It does not resolve the transient dynamics of the wakes, but can be used to simulate successive steady states of the farm.
PD is computed from the steady-state available powers.
Each turbine is controlled using the \textit{simple derating} operation mode of FLORIS.
This turbine mode curtails the power of the turbine to the allocated setpoint, with assumption that the thrust coefficient scales directly with power (i.e constant axial induction factor).

FAST.Farm is a dynamic simulator, which resolves the temporal dynamics of the wind farm, including the wake propagation. This is a more complex simulator than FLORIS, hence slower to execute, which enables us to increase the fidelity of our simulations.
The power tracking controller, in FAST.Farm simulations, for each turbine, relies on torque and pitch control, referred to as pitch Active Power Control (pitch APC) in
\cite{ahoActivePowerControl2016}.
Given a power reference for the turbine $P_i$, the rated generator speed $\omega_{g,\text{ref}}$ is set as:
\begin{equation}
    \omega_{g,\text{ref}} = \sqrt[3]{\frac{P_i}{\kappa}}
\end{equation}
where $\kappa$ is a constant depending on the turbine characteristics and the wind speed, used in maximum power point tracking control \cite{bossanyiDesignClosedLoop2000}.
The generator torque is then set as:
\begin{equation}
    T_g = \min\left(\kappa \omega_g^2, \frac{P_i}{\eta \omega_{g,\text{ref}}}\right)
\end{equation}

The blade pitch angle $\beta$ is then controlled using a Proportional-Integral (PI) controller to
ensure power tracking:
\begin{equation}
    \beta = K_P(P - P_i) + K_I \int (P - P_i) dt
\end{equation}
where $K_P$ and $K_I$ are the proportional and integral gains of the controller, respectively.

In all the cases simulated in this section, the turbine model used is the NREL 5MW reference turbine \cite{jonkmanDefinition5MWReference2009}, with a rotor diameter of $D = \SI{126}{\meter}$.

\subsection{Comparison of IPD applied to FLORIS and FAST.Farm}

\begin{figure*}[t]
    \centering
    \resizebox{\linewidth}{!}{\includegraphics{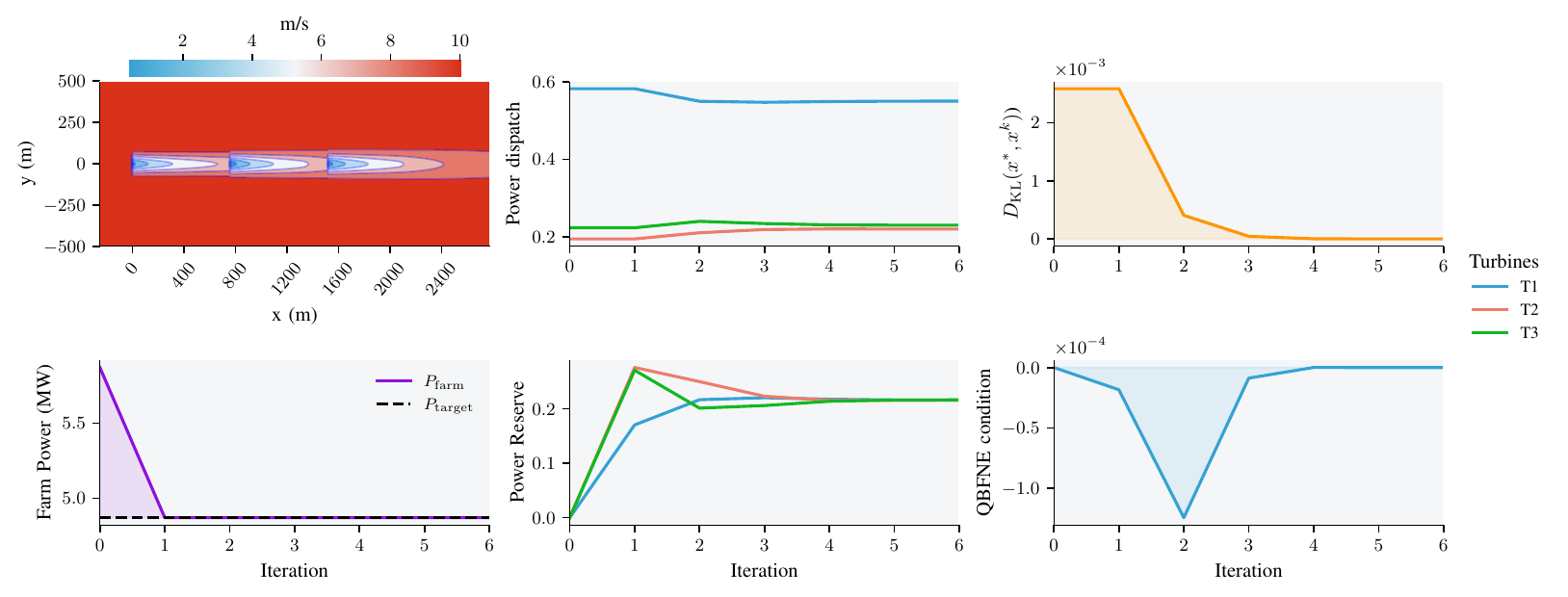}}
    \caption{IPD for a farm of 3 turbines in a row simulated on FLORIS, with an aligned wind of velocity \SI{10}{\meter\per\second} and a power target of \SI{1}{\mega\watt} below the greedy power.
    Top-left: wind field. Bottom-left: total farm power and target power. Top-center: power setpoints for each turbine. Bottom-center: power reserves of each turbine. Top-right: KL divergence to the fixed point (has to be non-increasing). Bottom-right: QBFNE condition (\cref{prop:coupled}, has to be negative)}
    \label{fig:dispatch_3}
\end{figure*}

\begin{figure*}[t]
    \centering
    \resizebox{\linewidth}{!}{\includegraphics{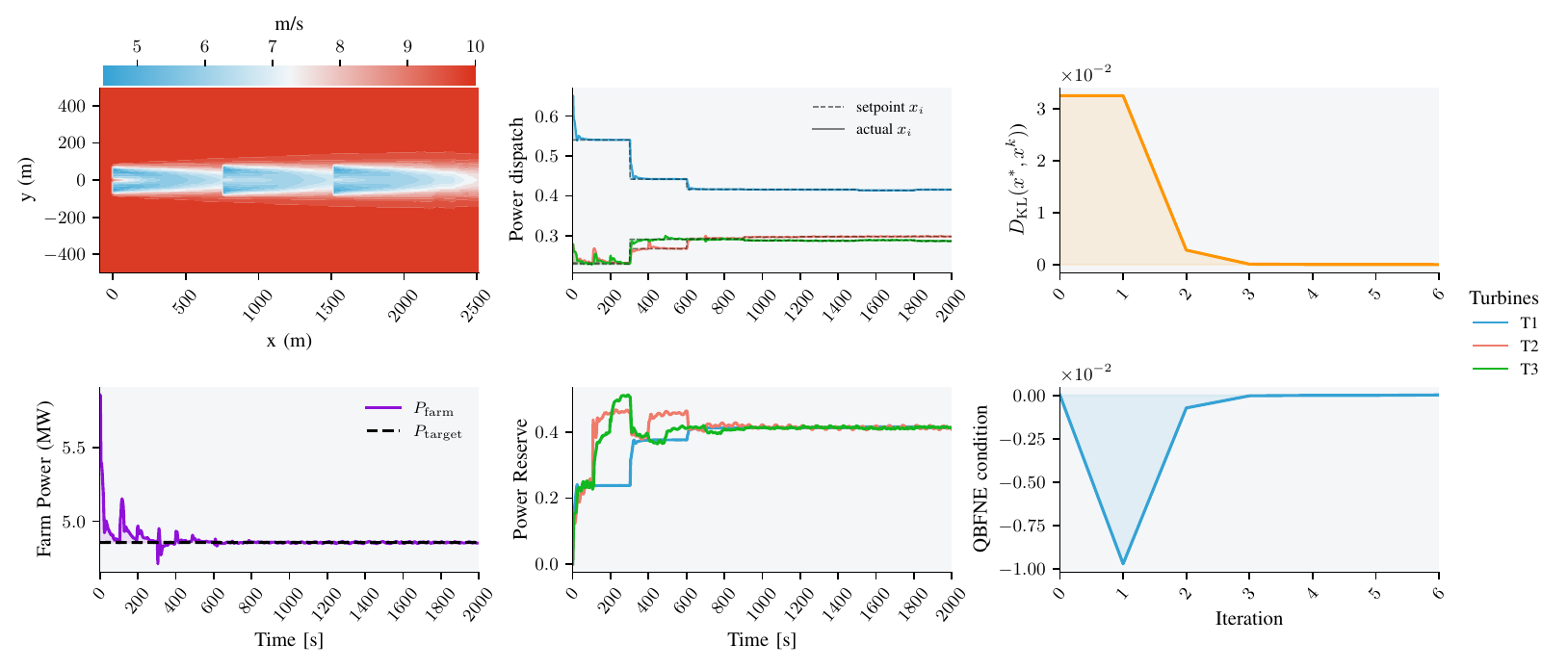}}
    \caption{IPD for a farm of 3 turbines in a row, with an aligned steady wind of velocity \SI{10}{\meter\per\second} and a power target of \SI{3}{\mega\watt} below the greedy power, simulated with FAST.Farm.
    Top-left: wind field. Bottom-left: total farm power and target power. Top-center: power setpoints for each turbine (dashed lines represent the actual power share realized). Bottom-center: power reserves of each turbine. Top-right: KL divergence to the fixed point (has to be non-increasing). Bottom-right: QBFNE condition (\cref{prop:coupled}, has to be negative)}
    \label{fig:dispatch_3_fast_farm}
\end{figure*}

We first consider the case of a farm with 3 turbines in a row, spaced by $6D$, subject to an aligned wind of speed \SI{10}{\meter\per\second}.  The farm is required to reduce its power by \SI{1}{\mega\watt}.
We will simulate this case using IPD both with FLORIS and FAST.Farm simulators.

In \autoref{fig:dispatch_3}, we present the result of IPD with FLORIS.
The farm-level target is reached after the first iteration of IPD (bottom-left figure), but due to the updated wake effects, the power reserves of the turbines are not equal after the first iteration (bottom-center figure).
The reserve of the first turbine is lower than the reserves of the second and third turbines, which have benefited from the decrease in power of the first turbine having weakened its wake.
This illustrates that applying PD only once does not achieve fair load sharing, in the presence of wake coupling between turbines, and justifies the use of IPD.
It takes 4 iterations of IPD for the power setpoints to converge to a fixed point (top-center), where the power reserves of all turbines are equal to 0.21 (bottom-center), thus achieving fair load sharing.
The first turbine is subject to the free-stream wind, and therefore has the highest available power.
Therefore, it is allocated the most power to produce by the IPD, almost 55\%, while the last two turbines are allocated around 22\% each.
The right column of \autoref{fig:dispatch_3} verifies the theoretical sufficient condition for convergence derived in \cref{prop:coupled}.
The KL divergence (top-right) to the fixed point is non-increasing, and the QBFNE condition \eqref{eq:QBFNE_conditions} (bottom-right) is negative at each iteration as expected.

In \autoref{fig:dispatch_3_fast_farm}, we also show results on the same case simulated on FAST.Farm.
The dispatched setpoints are updated every \SI{300}{\second}, so that the wakes have time to propagate and reach a steady state.
The results are very similar to the ones obtained on FLORIS, although the power levels and setpoints are slightly different due to the different models. This justifies using only the steady state simulator FLORIS for the remaining examples, which is less costly to compute.
Until \SI{100}{\second}, the power reserves are close to equal between the turbines after the first iteration. But then, the updated wake of turbine 1 reaches turbines 2 then 3 at \SI{200}{\second} and \SI{300}{\second} respectively, leading to an increase in their available power and therefore in their power reserves.
It finally takes 4 iterations of IPD to reach a fair load sharing solution, with power reserves of around 0.4 for each turbine. The power reserves reach a common value after 900 seconds.
The QBFNE condition is verified (negative) at each iteration, and the KL divergence to the fixed point is non-increasing.
Other simulations have been conducted with a more frequent update of the setpoints, also showing convergence after a few updates of the wakes.

\subsection{Examples on layouts with more turbines}

\begin{figure*}[t]
    \centering
    \resizebox{\linewidth}{!}{\includegraphics{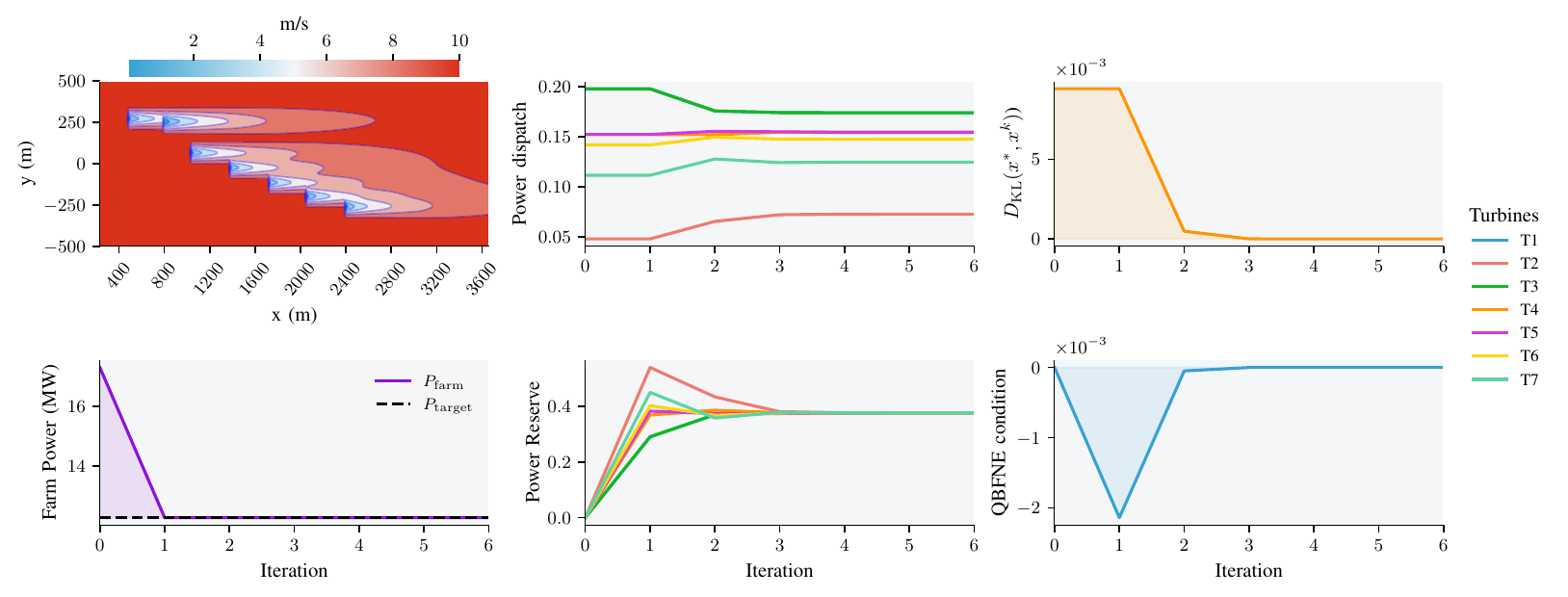}}
    \caption{IPD for the Ablaincourt SMV farm simulated on FLORIS, with an aligned wind of velocity \SI{10}{\meter\per\second} and a power target of \SI{5}{\mega\watt} below the greedy power.
    Top-left: wind field. Bottom-left: total farm power and target power. Top-center: power setpoints for each turbine. Bottom-center: power reserves of each turbine. Top-right: KL divergence to the fixed point (has to be non-increasing). Bottom-right: QBFNE condition (\cref{prop:coupled}, has to be negative)}
    \label{fig:dispatch_ablaincourt}
\end{figure*}

In \autoref{fig:dispatch_ablaincourt}, we consider the case of the existing Ablaincourt SMV onshore farm, composed of 7 turbines, studied in~\cite{ducLocalTurbulenceParameterization2019}.
In this case, most of the turbines are only partially waked, except the top-two turbines which form a row aligned with the wind, spaced by 2.5$D$.
The target power is set to \SI{5}{\mega\watt} below the greedy power.
After one application of PD, the power reserves are far from equal between the seven turbines, ranging from 0.29 to 0.53.
In only 3 iterations of IPD, the power setpoints converge to a fixed point and fair load sharing is achieved, with power reserves of around 0.38 for each turbine.
Once again the QBFNE condition \eqref{eq:QBFNE_conditions} is verified (negative) at each iteration, and the KL divergence to the fixed point is non-increasing, which is consistent with \cref{prop:coupled}.

\begin{figure*}[t]
    \centering
    \resizebox{\linewidth}{!}{\includegraphics{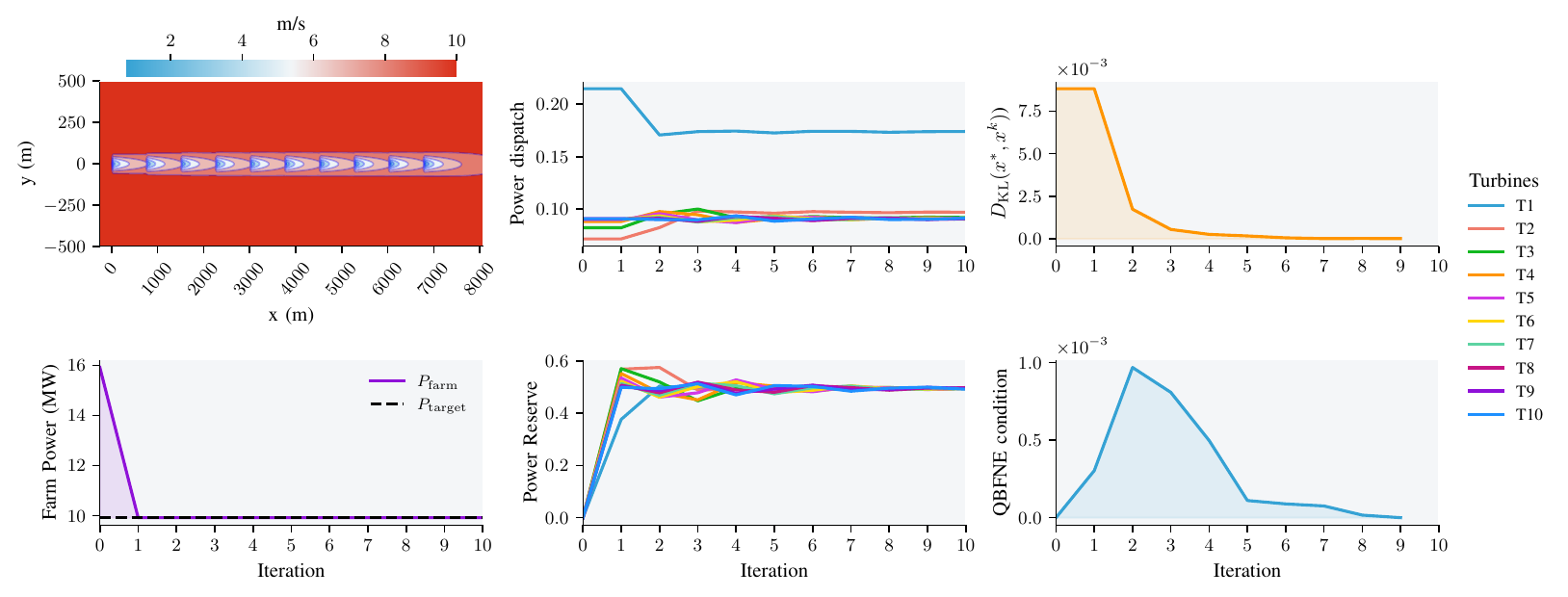}}
    \caption{IPD for a farm of 10 turbines in a row simulated on FLORIS, with an aligned wind of velocity \SI{10}{\meter\per\second} and a power target of \SI{6}{\mega\watt} below the greedy power.
    Top-left: wind field. Bottom-left: total farm power and target power. Top-center: power setpoints for each turbine. Bottom-center: power reserves of each turbine. Top-right: KL divergence to the fixed point (has to be non-increasing). Bottom-right: QBFNE condition (\cref{prop:coupled}, has to be negative)}
    \label{fig:dispatch_10}
\end{figure*}

We then show results on a case where the sufficient QBFNE condition on the dispatch operator is not verified, but where IPD still converges to a fair load sharing solution.
In \autoref{fig:dispatch_10}, we consider a farm of 10 turbines in a row, spaced by $6D$.
The power allocations and reserve converge in 8 iterations, leading to a fair load sharing solution.
In this case, we can see in the right column of \autoref{fig:dispatch_10} that the QBFNE condition is not satisfied (positive) at each iteration.
However, the KL divergence to the fixed point is still non-increasing, as explained in Remark~\ref{remark:2}.

In this case, the dynamics of the system are more complex because the row of turbines is longer.
The wake effects of the first turbine do not reach the last turbines of the row, but they are still coupled through the intermediate turbines and the dispatch scheme.
This creates some complex interactions: for example, a decrease in power of the first turbine leads to an increase in available power (and therefore in power setpoint) of the second turbine, which leads to a decrease in available power for the third turbine, and so on along the row.
The number of turbines in the row creates highly non-linear dynamics between the first and last turbines, which can lead to oscillations in the power setpoints and reserves, as observed in the bottom center of \autoref{fig:dispatch_10}.
This case illustrates a complex scenario of many fully waked turbines, which might not be very common in practice. They still display the robustness of IPD to converge to a fair load sharing solution, even when the sufficient condition is not verified (Remark~\ref{remark:2}).
In all cases shown in this section, applying PD only once does not achieve fair load sharing, contrary to beliefs in the literature, confirming that IPD is needed to reach a fixed point with equal power reserves.

\subsection{IPD under unsteady wind conditions}

\begin{figure}[h]
\centering
\resizebox{0.5\linewidth}{!}{\includegraphics{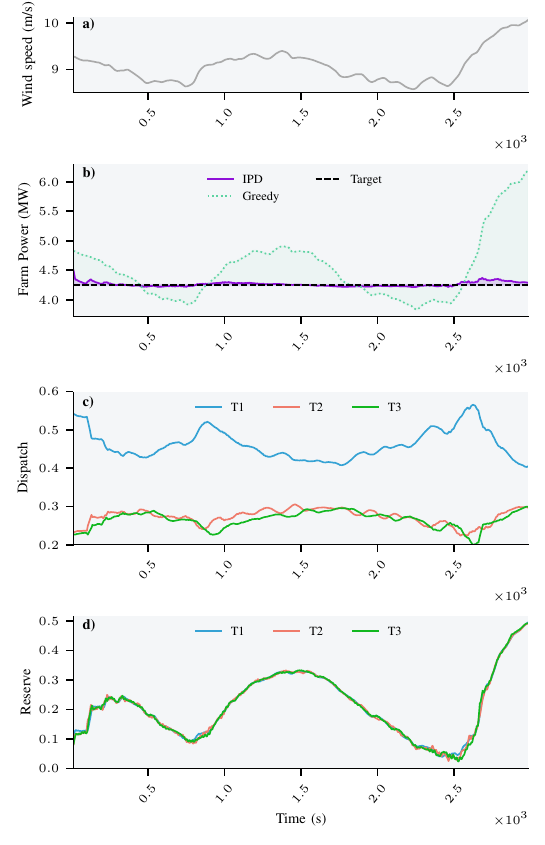}}
\caption{IPD on a 3-turbine row under time-varying wind speed (a), tracking a constant power target. (b) Farm power output under IPD and greedy operation. (c) Dispatch with IPD. (d) Corresponding power reserves}
\label{fig:varying_wind}
\end{figure}

We now investigate the behavior of IPD under time-varying wind inflow. To this end, we simulate in FAST.Farm a 3-turbine row subject to a time-varying wind speed. The wind speed signal is extracted from measurements conducted at the Ablaincourt SMV wind farm~\cite{ducLocalTurbulenceParameterization2019}. The corresponding time series is shown in \autoref{fig:varying_wind}a.

In \autoref{fig:varying_wind}b, the farm power target (black dashed line) is constant and set to \SI{4.25}{\mega\watt}. The farm output under IPD (purple line) closely tracks this target, with an RMS tracking error of \SI{0.05}{\mega\watt}.

The dispatched power shares are shown in \autoref{fig:varying_wind}c. The IPD update frequency is set to \SI{3}{\second} to capture wind fluctuations. The convergence process is visible during the first \SI{400}{\second}: the allocation of the upstream turbine decreases in successive steps, enabling increased contributions from downstream turbines as the wake propagates. Beyond this transient phase, the dispatch evolution becomes strongly correlated with the wind inflow variations.

As shown in \autoref{fig:varying_wind}d, the power reserves rapidly converge to a common value in \SI{100}{\second}.
Then, the equal reserve evolves with the successive wake adjustments induced by the dispatch updates. This is visible by the steps in the first \SI{400}{\second}.
For the rest of the simulation, the shared reserve value varies, mainly in relation to the incoming wind speed evolution rather.

Due to wind variability, the instantaneous greedy power occasionally drops below the target (\autoref{fig:varying_wind}b). However, the farm controlled with IPD is still able to maintain the required power. It is visible in figure d) that the common reserve obtained with IPD never drops to zero during those momentary drops in wind speed.
No turbine saturates, therefore enabling the farm to produce the desired power.
This example displays the robustness of fair load sharing dispatch to wind fluctuations and uncertainties.

\subsection{Max-min fairness of the IPD solution}

Then, we numerically evaluate if the fair load sharing solution found by applying IPD also achieves max-min fairness.
For that, we solve the optimization problem~\eqref{eq:critere_optim} directly using two optimization methods: a differential evolution algorithm~\cite{stornDifferentialEvolutionSimple1997} and COBYQA with multiple random starts, which is a derivative-free, trust-region Sequential Quadratic Programming (SQP) approach~\cite{ragonneauModelbasedDerivativefreeOptimization2023}.
In \cref{tab:res_optim}, we compare the minimum power reserve obtained by the two optimization methods with the power reserve obtained by the IPD solution, for the different cases shown in this section.
The \cref{tab:res_optim} also displays the number of function evaluations the optimization algorithm was allowed to perform.
For the differential evolution, this number is computed as the product of the population size, the number of generations and the number of turbines in the farm.
These results show that the solution found by the optimization methods is equal, or only slightly better, than the one obtained with IPD. This suggests that the IPD solution is a very good approximation of a max-min fairness solution.
Moreover, the number of function calls needed by the optimization to converge grows with the number of turbines. For differential evolution, depending on the layout considered, it ranges from $10^3$ to $10^5$ simulator calls, which can be computationally costly if performed on multiple different wind conditions. On the other hand, IPD only requires few iterations to converge.

\begin{table}[htbp]
\centering
\footnotesize
\setlength{\tabcolsep}{3pt}
\renewcommand{\arraystretch}{1.35}

\begin{tabular}{@{}p{1.7cm}ccccc@{}}
\toprule
\textbf{Case} &
\textbf{IPD} &
\textbf{DE} &
\textbf{COBYQA} &
\textbf{DE eval.} &
\textbf{COBYQA eval.} \\
\midrule

\makecell[l]{3-turbine row}
& 0.2163
& 0.2165
& 0.2165
& $3\times10^3$
& 70 \\

\makecell[l]{Ablaincourt\\(7 turbines)}
& 0.3760
& 0.3762
& 0.3760
& $3.5\times10^4$
& 3781 \\

\makecell[l]{10-turbine row}
& 0.4955
& 0.4955
& 0.4943
& $3\times10^5$
& 15677 \\

\bottomrule
\end{tabular}

\vspace{0.3cm}

\caption{Comparison of the max-min reserve obtained using differential evolution (DE) and multistart COBYQA against IPD.}
\label{tab:res_optim}

\end{table}

\FloatBarrier

%% file: yaw_augmented.tex
% !TEX root = main.tex
In~\cite{tamaroRobustActivePower2025a}, the authors propose a yaw-augmented dispatch strategy. The power setpoints and yaw angles of each turbine are jointly optimized for the max-min fairness objective.
Using wake steering can increase the available power of downstream turbines, thus improving the overall power tracking performance of the farm.
However, this doubles the number of variables to optimize.
In this section, we show that IPD is also very effective in this extended setting to simplify the optimization problem.

The yaw-augmented dispatch problem can be formulated as follows:
\begin{equation}
\label{eq:maxmin_yaw}
\max_{x, \gamma} \min_i r_i(x,\gamma) \;\; \text{s.t.} \;\; x\in\Delta,\; x_i \leq \frac{P_{a,i}(x,\gamma)}{\ptarg}
\end{equation}
where  $\gamma = (\gamma_i)_{i=1}^N$ is the yaw angle vector. The available power of each turbine now also depends on the yaw angles of the turbines, as well as their power setpoints.
In order to take into account misalignment effects, the available power is computed using a cosine-loss correction~\cite{gebraadWindPlantPower2016}:
\begin{equation}\label{eq:Pa_def_yaw}
P_{a,i}(x,\gamma)= \frac{1}{2} \rho A C_{p,\max} u_i(x,\gamma)^3 \cos(\gamma_i)^{P_p}
\end{equation}
In FLORIS, this exponent was calibrated on high fidelity LES simulations~\cite{gebraadWindPlantPower2016} and is by default $P_p = 1.88$.

Considering that IPD converges to a close-to-optimal solution in the coupled setting, we can consider this simplified optimization problem:

\begin{equation}
\label{eq:yaw_IPD_problem}
\max_{\gamma} \min_i r_i^\star(\gamma) 
\end{equation}
where $r_i^\star(\gamma)$ is the power reserve of turbine $i$ at the converged dispatch of IPD for a given yaw angle vector $\gamma$.
This reduces the number of optimization variables from $2N$ to $N$, with $N$ the number of turbines, as we only need to optimize over the yaw angles, while the power setpoints are directly given by IPD.

\begin{figure}[htbp]
    \centering
    \resizebox{0.58\linewidth}{!}{\includegraphics{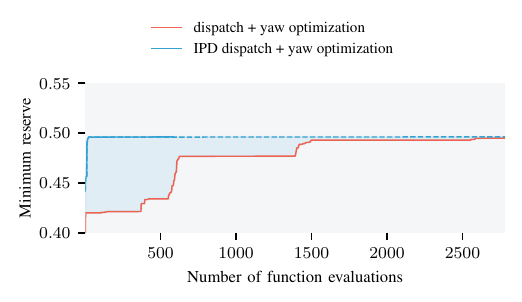}}
    \caption{Comparison of yaw-augmented dispatch optimization for max-min fairness using COBYQA, with (blue) and without (red) IPD, on a 5-turbine row.
            The evolution of the minimum power reserve in the farm is plotted as a function of the number of function evaluations by the algorithm.}\label{fig:yaw_optimization_comparison}
\end{figure}

To illustrate this, we simulate with FLORIS a row of 5 turbines spaced by 6$D$, under a \SI{10}{\meter\per\second} aligned wind, required to reduce its power by \SI{3}{\mega\watt}.
As previously, we use COBYQA~\cite{ragonneauModelbasedDerivativefreeOptimization2023} with random multi-starts to solve the optimization problems~\eqref{eq:maxmin_yaw} and \eqref{eq:yaw_IPD_problem}.

Solving problem~\eqref{eq:maxmin_yaw} requires optimizing over both $\gamma$ and $x$ (9 variables, as the first 4 dispatch fractions constrain the 5th), while problem~\eqref{eq:yaw_IPD_problem} requires optimizing only over $\gamma$ (5 variables).
In \autoref{fig:yaw_optimization_comparison}, we show the evolution of the minimum reserve as a function of the number of function evaluations (i.e. simulator calls) by the optimization algorithm.
The optimization of both yaw and setpoints took 3573 function calls to converge. It reached a sub-optimal solution of slightly uneven reserves between turbines, with a minimum reserve of 0.4950.
In comparison, the optimization of yaw angles with IPD converged after 818 function calls, and reached perfect load sharing with a common reserve of 0.4962. This represents a 77\% reduction in number of simulator calls.
This shows the potential of IPD to greatly speed-up the optimization of yaw-augmented fair dispatch, by halving the number of optimization variables.

\begin{figure}
    \centering
    \resizebox{0.58\linewidth}{!}{\includegraphics{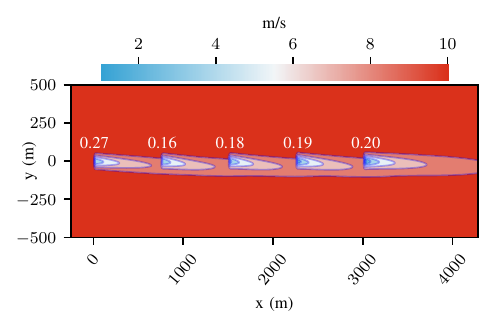}}
    \vspace{-0.2cm}
    \caption{Wind field for a 5 turbine row with yaw angles optimized for max-min fairness of power reserves, using IPD dispatch (white values above each turbine).}
    \label{fig:optimized_wind_field}
\end{figure}

In \autoref{fig:optimized_wind_field}, we show the wind field obtained with the optimized yaw and dispatch (IPD and yaw optimization). The optimized yaw are $\gamma=(21, 22, 19, 13, 0)^\circ$. The dispatch obtained by applying IPD is denoted in white above each turbine.
The max-min reserve without enabling yaw steering is 0.448. Therefore, the yaw optimization leads in this case to an increase of 11\% in reserve.

%% file: conclusion.tex
% !TEX root = main.tex
In this article, we have studied how to achieve power tracking with wind farms, by dispatching the target power with a fair repartition of power reserves.
This helps with robustness to uncertainties and changing wind conditions.
We studied the fairness properties of proportional dispatch, in the presence or not of wake-coupling between turbines. 

Without wake effects, proportional dispatch achieves both max-min fairness and fair-load sharing. 
Then, with wake effects, we introduced an iterative proportional dispatch (IPD), in order to consider the wake updates due to each change in power setpoints. We have proven convergence of this process to a fair load sharing solution, under sufficient conditions. The presented numerical experiments demonstrate that IPD indeed achieves fair load sharing, and closely approaches max-min fairness.
These experiments were conducted on both steady-state and dynamic simulators, on a variety of wind farm layouts.
Finally, we considered the extended problem of fair dispatch combined with yaw wake steering. IPD allowed to halve the number of variables to optimize, thus reducing drastically the computational cost.

Future work will need to make the convergence analysis more interpretable, by expressing the conditions with physical properties of the system (such as wake parameters).

%% file: appendixA.tex
% !TEX root = main.tex
\begin{myproof}

Let us define $\vect{P_a} \in \R^N$ the vector of available powers, and for any dispatch $\vect{P} \in \R^N$, we define $\vect{r} \in \R^N$ the vector of reserves with the $i$-th element expressed as: \( r_i = 1 - \frac{P_i}{P_{a,i}} \).

According to the Hölder inequality, we have:
\[
|\vect{P_a}^\top (\1-\vect{r})| \leq \|\vect{P_a}\|_1 \|\1-\vect{r}\|_\infty
\]
where \( \|\vect{P_a}\|_1 = \sum_{i=1}^N |P_{a,i}| = \sum_{i=1}^N P_{a,i} \), and \( \|\1-\vect{r}\|_\infty = \max_i |1-r_i| = \max_i \frac{P_i}{P_{a,i}} \).

Yet, for our problem,
\[
\vect{P_a}^\top (\1-\vect{r}) = \sum_{i=1}^N P_i =  \ptarg.
\]
Therefore,
\[
\ptarg \leq \|\vect{P_a}\|_1 \|\1-\vect{r}\|_\infty
\]
This implies that:
\[
 \frac{\ptarg}{\|\vect{P_a}\|_1} \leq \|\1-\vect{r}\|_\infty 
\]
which gives
\[
1 - \frac{\ptarg}{\|\vect{P_a}\|_1} \geq \min_i r_i
\]
And this bound is reached by setting \( P_i^\star = \frac{P_{a,i}}{\sum_{j=1}^N P_{a,j}} \ptarg \) for all \( i \) which is the definition of PD \eqref{eq:PD_def}.

Additionally, using this PD equalizes the reserves for all turbines to $r^\star$:
\[ r^\star = 1 - \frac{ P_i^\star}{P_{a,i}} = 1 - \frac{ \ptarg}{\sum_{j=1}^N P_{a,j}} \]

Therefore, max-min fairness is achieved by proportional dispatch, which equalizes the reserves at the maximal possible value \( r^\star \).
\end{myproof}

%% file: appendixB.tex
% !TEX root = main.tex

\begin{definition}[Bregman and KL divergence] 
Let $\psi:\mathbb{R}^N\to\mathbb{R}$ be a strictly convex and differentiable function, 
and let $x,y\in\mathbb{R}^N$. The Bregman divergence associated with $\psi$ is defined as
\[
D_\psi(x,y)
=
\psi(x)-\psi(y)-\langle\nabla\psi(y),x-y\rangle,
\]
where $\langle\cdot,\cdot\rangle$ denotes the standard inner product in $\mathbb{R}^N$ 
and $\nabla\psi(y)$ is the gradient of $\psi$ evaluated at $y$.

In particular, let $x,y\in\Delta$ and consider the negative entropy function
\[
\psi(x)=\sum_{i=1}^N x_i\log x_i.
\]
The associated Bregman divergence is the Kullback-Leibler (KL) divergence~\eqref{def:KL}.
\end{definition} 

We will use in the proof the following standard properties of the KL divergence~\cite{coverElementsInformationTheory2005}:
For all $x,y \in \Delta$:
\begin{itemize}
  \item non-negativity: $D_{\text{KL}}(x,y) \geq 0$, with equality if and only if $x=y$.
  \item Three-point identity: For all $x,y,z \in \Delta$, we have: 
  \begin{align*}
    D_{\text{KL}}(x,z) &= D_{\text{KL}}(x,y) + D_{\text{KL}}(y,z) \\
    &\quad + \langle x - y, \nabla \psi(y) - \nabla \psi(z) \rangle
  \end{align*}
\end{itemize}

We next present the proof of \cref{prop:coupled}.

\begin{myproof}

\paragraph*{Existence of a fixed point}

$g$ is continuous, as it is a composition of continuous functions. It maps the compact convex set $\Delta$ to itself. Thus, by Brouwer's fixed point theorem, $g$ has at least one fixed point in $\Delta$.

\paragraph*{Convergence of the iterates to a fixed point}
We aim at showing that the sequence $x_k$ has a limit which is $x^\star$. To do so, we are interested in the `distance' between the iterates and a fixed point $x^\star$, measured by $V^k=D_{\text{KL}}(x^\star, x^k)$. 
We want to show that $V^k$ is a non-increasing sequence.

Using the three-point identity of the KL divergence, we have:
\begin{multline}\label{eq:vk}
  V^k = V^{k+1} + D_{\text{KL}}(x^{k+1}, x^k) \\
  + \langle x^\star - x^{k+1}, \nabla \psi(x^{k+1}) - \nabla \psi(x^k) \rangle
\end{multline}
The QBFNE condition on $g$ gives 
\begin{equation}
\langle x^\star - x^{k+1}, \nabla \psi(x^{k+1}) - \nabla \psi(x^k) \rangle \geq 0
\end{equation}
This gives $V^k \geq V^{k+1}$. It is a sufficient condition, however even if the QBFNE condition is not satisfied, \eqref{eq:vk} can still give $V^k \geq V^{k+1}$ thanks to the positive term $D_{\text{KL}}(x^{k+1}, x^k)$. This scenario is observed in the last two examples of \autoref{sec:numerical}.

Therefore,
$V^k$ is a non-increasing sequence bounded below by $0$, so it converges to some limit $V^\infty \geq 0$.
This implies that when $k \to \infty$, we have $V^k - V^{k+1} \to 0$. The relation~\eqref{eq:vk}, and the non-negativity of the KL divergence, imply that
\begin{equation}\label{eq:kl_limit}
    \lim_{k \to \infty} D_{\text{KL}}(x^{k+1}, x^k) = 0
\end{equation}
Because the iterates $\{x^k\}$ belong to the compact set $\Delta$, Bolzano-Weierstrass theorem guarantees the existence of a convergent subsequence $\{x^{k_j}\}$ that converges to some limit point $x^\star \in \Delta$.
By the continuity of $g$ and $D_{\text{KL}}$, and the relation \eqref{eq:kl_limit}, we have
\begin{equation}
    D_{\text{KL}}(g(x^\star), x^\star) = \lim_{j \to \infty} D_{\text{KL}}(x^{k_j+1}, x^{k_j}) = 0
\end{equation}
By property of the KL divergence, we get $g(x^\star) = x^\star$. Thus, every limit point of the sequence is a fixed point of $g$.

Finally, we show that the entire sequence converges to a unique point. 
Since the subsequence $D_{\text{KL}}(x^\star, x^{k_j})$ converges to $0$, the entire sequence ${V^k}=\{D_{\text{KL}}(x^\star, x^k)\}$ must converge to $V^\infty=0$. 
Therefore by non-negativity, $\lim_{k \to \infty} x^k = x^\star$, and the iterates converge to a fixed point of $g$.

\end{myproof}